\documentclass[twocolumn,showpacs,superscriptaddress,preprintnumbers,amsmath,amssymb]{revtex4}

\usepackage{graphicx}
\usepackage{color}
\usepackage{dcolumn}
\usepackage{bm}

\begin{document}


\title{Hard ellipses: Equation of state, structure and self-diffusion}

\author{Wen-Sheng Xu}
\affiliation{State Key Laboratory of Polymer Physics and Chemistry, Changchun Institute of Applied Chemistry, Chinese Academy of Sciences, Changchun 130022, People's Republic of China}
\affiliation{James Franck Institute, The University of Chicago, Chicago, Illinois 60637, USA}

\author{Yan-Wei Li}
\author{Zhao-Yan Sun}
\email{zysun@ciac.jl.cn.}
\author{Li-Jia An}
\email{ljan@ciac.jl.cn.}
\affiliation{State Key Laboratory of Polymer Physics and Chemistry, Changchun Institute of Applied Chemistry, Chinese Academy of Sciences, Changchun 130022, People's Republic of China}



\date{\today}

\begin{abstract}
Despite their fundamental and practical interest, the physical properties of hard ellipses remain largely unknown. In this paper, we present an event-driven molecular dynamics study for hard ellipses and assess the effects of aspect ratio and area fraction on their physical properties. For state points in the plane of aspect ratio ($1\leq k \leq9$) and area fraction ($0.01\leq \phi \leq0.8$), we identify three different phases, including isotropic, plastic and nematic states. We analyze in detail the thermodynamic, structural and self-diffusive properties in the formed various phases of hard ellipses. The equation of state (EOS) is shown for a wide range of aspect ratios and is compared with the scaled particle theory (SPT) for the isotropic states. We find that SPT provides a good description of the EOS for the isotropic phase of hard ellipses. At large fixed $\phi$, the reduced pressure $p$ increases with $k$ in both the isotropic and the plastic phases and, interestingly, its dependence on $k$ is rather weak in the nematic phase. We rationalize the thermodynamics of hard ellipses in terms of particle motions. The static structures of hard ellipses are then investigated both positionally and orientationally in the different phases. The plastic crystal is shown to form for aspect ratios up to $k=1.4$, while appearance of the stable nematic phase starts approximately at $k=3$. We quantitatively determine the locations of the isotropic-plastic (I-P) transition and the isotropic-nematic (I-N) transition by analyzing the bond-orientation correlations and the angular correlations, respectively. As expected, the I-P transition point is found to increase with $k$, while a larger $k$ leads to a smaller area fraction where the I-N transition takes place. Moreover, our simulations strongly support that the two-dimensional nematic phase in hard ellipses has only quasi-long-range orientational order. The self-diffusion of hard ellipses is further explored and connections are revealed between the structure and the self-diffusion. We discuss the relevance of our results to the glass transition in hard ellipses. Finally, the results of the isodiffusivity lines are evaluated for hard ellipses and we discuss the effect of spatial dimension on the diffusive dynamics of hard ellipsoidal particles.
\end{abstract}


\maketitle

\section{Introduction}

Hard-particle systems in general offer a great opportunity to understand the mechanism of a variety of physical phenomena, including crystal nucleation~\cite{Auer, Gasser}, glass transition~\cite{Pusey, Kegel, Weeks, Patrick1, Patrick2} and jamming~\cite{Torquato, Patrick3}. The spherical particles, such as hard spheres and hard disks, have been studied intensively during the past decades. In particular, for a system of hard spheres, an entropy-driven liquid-crystal transition occurs at sufficiently high density~\cite{Alder} and a hard-sphere glass can also be formed above the glass transition point if crystallization is avoided~\cite{Zaccarelli}. On the other hand, non-spherical particles play a fundamental role in the physics of molecular liquids and attract growing interest in recent years because of their usefulness in forming functional structures. From the theoretical point of view, it is interesting to include the rotational degrees of freedom for the particles and investigate the physical behavior of a system composed of anisotropic particles. Such study uncovers new types of behavior and holds potential for providing new insights that cannot be obtained from studies of spherical objects~\cite{Frenkel, DeMichele, Donev1, Donev2, Mailman, Letz, Zheng}. From the practical point of view, the constituent particles often have non-spherical shapes in real materials such as laponite~\cite{Ruzicka}, and the building blocks with anisotropic shape can provide a powerful candidate for the assembly of particular targeted structures~\cite{Glotzer, Damasceno}. Thus, exploration of systems composed of anisotropic particles is also valuable for materials design. In this paper, we present an event-driven molecular dynamics (EDMD) study of hard ellipses. A system of hard ellipses can be regraded as one of the simplest models of two-dimensional (2D) anisotropic particles and its phase behavior is very rich. Yet, the thermodynamic, structural and dynamical properties of hard ellipses remain largely unknown.

The phase transitions, in particular, the isotropic-nematic (I-N) transition, of hard ellipses have been studied by computer simulations and various theoretical approaches. In Vieillard-Baron's pioneering Monte Carlo (MC) simulations in $NVT$ ensemble (i.e., the particle number $N$, the volume $V$ and the termperature $T$ are constant)~\cite{VieillardBaron}, three different phases, including isotropic liquid, nematic liquid crystal and solid, have been identified in a hard-ellipse system with an aspect ratio of $k=a/b=6$, where $a$ and $b$ denote semi-major and semi-minor axes. Vieillard-Baron observed that both I-N and nematic-solid (N-S) transitions are of first order, as evidenced by a small discontinuity shown in the equation of state (EOS, i.e., density-dependence of the reduced pressure $p$). However, computer simulations show that the I-N transition in a 2D fluid of hard needles~\cite{Frenkel1} and long hard rods~\cite{Bates}, which in principle share the similar physics with elongated hard ellipses, is continuous rather than first-order. Moreover, the $NPT$ (i.e., the particle number $N$, the pressure $P$ and the termperature $T$ are constant) MC simulations of Cuesta and Frenkel~\cite{Frenkel2} indicate that while there is no stable nematic phase in hard ellipses for $k=2$, the I-N transition is of first order for $k=4$ and continuous via disclination unbinding for $k=6$. Hence, it appears that the nature of the I-N transition depends on the aspect ratio and that the phase diagram possesses a tricritical point at which the transition changes from first-order to continuous. Thus, on the simulation side, a definite conclusion has not been reached on the phase transitions of hard ellipses. This issue is also contentious on the theoretical side. Ward and Lado~\cite{Ward} found based on the Percus-Yevick (PY) integral equation theory that the I-N transition does not take place in a fluid of hard ellipses although orientational ordering is allowed. While, the density-functional theory (DFT)~\cite{Cuesta1} and the effective-liquid approach~\cite{Cuesta2} indicate that the I-N transition is continuous for hard ellipses with arbitrary aspect ratios. The other forms of DFT~\cite{Ferreira, Varga}, the Onsager theory-based approach~\cite{Schoot},  and the scaled particle theory (SPT)~\cite{Schlacken} predict the existence of the I-N transition in hard ellipses, but its location differs from these theoretical predictions. Therefore, a clear understanding of the phase transitions in hard ellipses is still lacking. Here, it should also be mentioned that the phase behavior of binary hard ellipses was recently investigated by a scaled particle DFT~\cite{MR} and a weighted DFT~\cite{Moradi}.

Another important issue concerns the orientational order in the 2D nematic liquid crystals. Straley has shown~\cite{Straley} that true long-range order (LRO) cannot exist in a 2D nematic phase if the particles interact via a separable potential, but the existence of true LRO cannot be excluded if the potential is not separable into positional and orientational parts. In fact, it has been demonstrated by computer simulations~\cite{Frenkel1, Bates} that a general property of a 2D nematic phase is the lack of true LRO, although earlier simulations~\cite{Tobochnik} also show that the system with the nonseparable potential exhibits true LRO. For this reason, the systems that lack true LRO are usually referred to as having quasi-LRO. In the elastic continuum theory~\cite{DeGennes}, quasi-LRO is expected to appear in a 2D nematic phase if the free energy associated with collective fluctuations in the particle orientations can be expressed as
\begin{equation}
F=\frac{1}{2}\int K(\nabla\theta(\mathbf{r}))^{2}d\mathbf{r},
\end{equation}
where $\theta(\mathbf{r})$ characterizes the orientation at position $\mathbf{r}$ with respect to a fixed axis and $K$ is the 2D Frank's elastic constant~\cite{footnote1}. Based on Eq. (1), one can derive that the amplitude of the orientational fluctuations diverges with the system size in a logarithmic manner,
\begin{equation}
<\theta^{2}>\sim\frac{k_{B}T}{4\pi K}\ln N,
\end{equation}
where $<\cdot\cdot\cdot>$ denotes the ensemble average, $k_{B}$ is Boltzmann's constant, $T$ is the absolute temperature and $N$ is the particle number. As a consequence, both the 2D nematic order parameter $q$ and the angular correlation function $g_{2l}$ decay algebraically, i.e.,
\begin{equation}
q=<\cos(2\theta)>\sim N^{-k_{B}T/2\pi K},
\end{equation}
\begin{equation}
g_{2l}(r)=<\cos(2l[\theta(r)-\theta(0)])>\sim r^{-2l^{2}k_{B}T/\pi K},
\end{equation}
where $l$ is a positive integer. Equations (3) and (4) imply that a 2D nematic phase with quasi-LRO can be characterized by a vanishing orientational order parameter in the thermodynamic limit and a power-law decay of the angular correlation function. If the transition from a 2D nematic phase with quasi-LRO to an isotropic phase proceeds via the Kosterlitz-Thouless (KT) disclination unbinding mechanism~\cite{KT}, it should occur at a universal value of the renormalized Frank's constant, $\pi K_{c}/8k_{B}T=1$~\cite{Frenkel1}. As pointed out in Ref.~\cite{Frenkel1}, a different mechanism is also possible, but there is no stable nematic phase at values of the Frank's constant below $K_{c}$. Thus, one can quantitatively identify the location where the stable nematic phase starts to form by monitoring the system size dependence of the nematic order parameter or from the spatial correlation of the particle orientations. We note that the algebraic decay of the angular correlation function has been evidenced by simulations of 2D hard needles~\cite{Frenkel1}, 2D hard rods~\cite{Bates}, hard ellipses~\cite{Davatolhagh}, and recently by experiments of quasi-2D suspensions of hard ellipsoids~\cite{Zheng1}.

In contrast to the efforts made on understanding the phase behavior and the structure of hard ellipses, little attention has been paid on their dynamics despite its fundamental and technological importance. The dynamical properties of hard needles have been studied by molecular dynamics (MD) simulations about three decades ago~\cite{Frenkel3}. The diffusive process of a single hard ellipsoid confined in two dimensions was measured just several years ago in experiments~\cite{Han1, Han2}. In particular, the diffusion of quasi-2D suspensions of hard ellipsoids with $k\approx9$ has been explored very recently by video-microscopy experiments~\cite{Zheng1} and the enhancement of the translational diffusion with respect to the rotational diffusion has been rationalized in terms of the formation of unstable nematic-like regions with an average lifetime that exceeds the characteristic time of diffusion in a recent MC study~\cite{Davatolhagh}. In spite of the aforementioned progress, our knowledge of the dynamics of hard ellipses is far from complete. To our knowledge, there are no available MD results concerning the self-diffusion of hard ellipses so far.

In this work, we analyze in detail the thermodynamic, structural and self-diffusive properties in the various phases of hard ellipses, with the aim of examining the effects of aspect ratio and area fraction and exploring connections between different properties. To this end, we perform EDMD simulations for hard ellipses over a wide range of aspect ratios and area fractions, covering state points including the isotropic, plastic and nematic phases. We present the EOS for a series of $k$ values and compare the simulation results with a recent prediction of the SPT in the isotropic phase. We find that SPT well describes the isotropic branch of the EOS in hard ellipses. Dependence of $k$ on the reduced pressure $p$ is explored in the various phases. At large fixed area fractions, $p$ increases with $k$ in both the isotropic and the plastic phases, and interestingly, its dependence on $k$ is rather weak in the nematic phase. We rationalize the thermodynamics of hard ellipses in terms of particle motions, which show different features in different phases. The static structures, both positionally and orientationally, are then investigated in the various phases. Our simulations show that the plastic crystal forms for aspect ratios up to $k=1.4$, while appearance of the stable nematic phase starts approximately at $k=3$ for the studied density range. The locations of the isotropic-plastic (I-P) transition and the I-N transition are determined by analyzing the bond-orientation correlation functions and the angular correlation functions, respectively. As expected, the I-P transition point is found to increase with $k$, while a larger $k$ leads to a smaller area fraction where the I-N transition takes place. In addition, our results strongly support that the 2D nematic phase in hard ellipses has only quasi-LRO. We further explore the self-diffusion of hard ellipses. We find that a phase transition in the translational degrees of freedom is indeed reflected by an enhancement in the rotational diffusion and vice versa. Hence, a clear connection is revealed between the structure and the self-diffusion. We discuss the relevance of our results to the glass transition in hard ellipses, which is a subject of recent experimental investigation~\cite{Zheng, Weeks1}. We finally provide the results of the isodiffusivity lines for hard ellipses and discuss the effect of spatial dimension on the diffusive dynamics of hard ellipsoidal particles by comparing results of hard ellipses and uniaxial hard ellipsoids~\cite{DeMichele}.

The paper is organized as follows. In Sec. II, we specify the details of the simulation and the model system used in this work. Thermodynamic, structural and self-diffusive properties of hard ellipses are presented and discussed in Sec. III. A summary of our findings is given in Sec. IV. In the Appendix, we add the main ingredients for implementing an EDMD simulation for hard ellipses and a brief discussion on the effect of the system size.

\section{Simulation details}

We perform extensive EDMD simulations of hard ellipses in a square box under periodic boundary conditions in $NVT$ ensemble. The main ingredients for implementing an EDMD simulation for hard ellipses are described in the Appendix A. In a simulation, we fix the particle number $N$ and the area of the simulation box $V=L^{2}$ with $L$ the box dimension. In order to explore the properties in different phases, we investigate two sets of aspect ratios, i.e., $k=1-2$ with an interval of $0.1$ and $k=3-9$ with an interval of $1$, and a wide range of area fractions ranging from $\phi=N\pi ab/V=0.01$ to $0.8$. At sufficiently high densities, a nematic phase (i.e., particles have their centers of mass at random but exhibit some long-range orientational order) is expected to form for large $k$ values while a plastic phasealso known as the rotator phase, in which particles are oriented at random but their centers of mass have some ordered arrangement) can form for small $k$ values. All the particles have the same mass $m$ and the same moment of inertia $I$. Both $m$ and $I$ are set to be unity for convenience in our simulations. In principle, ellipses with the same mass but different aspect ratios should have different moment of inertia. In our simulations, the function of $m$ and $I$ (together with the particle velocities) is to determine the temperature $T$ of the system via equipartition theorem.  As we work on the athermal hard-particle system, the temperature only affects the time scale of the system (see the time unit below). Since we will report all results in reduced units, we expect that the results correctly capture the trends of static and dynamic quantities in variation with area fraction and aspect ratio. This is indeed strongly supported by the fact that our results for the static properties (such as the reduced pressure) are consistent with previously available MC simulations~\cite{Frenkel2}, where neither $m$ nor $I$ is relevant, suggesting that the general trends in this work are essentially unaffected by the choice of $m$ and $I$ once the results are given in reduced units. The temperature $T$ is irrelevant for athermal systems and remains constant due to the conservation of the total kinetic energy. We set $k_{B}T$ to be unity in this work. Note that $T$ contains both translational and rotational parts in the case of hard ellipses and that the separate temperature in each part fluctuates during the simulation. Length, time and pressure are expressed in units of $2b$, $\sqrt{4mb^{2}/k_{B}T}$ and $k_{B}T/4b^{2}$. In the following, we report results for $N=500$. However, as can be seen from Eq. (3), it is also useful to study the effect of the system size especially in the nematic phase. We thus also perform simulations of hard ellipses with $N=200$ and $800$ for selected state points, and a brief discussion on the effect of the system size will be shown in the Appendix B.

In order to generate the starting configuration of a hard-ellipse system with specific $k$ and $\phi$, we used the Lubachevsky-Stillinger (LS) compression algorithm~\cite{LS1, LS2}, in which the particles collide elastically and expand uniformly at a certain growth rate. LS algorithm was initially designed for hard spherical particles and recently it has been generalized to the case of hard ellipsoidal particles~\cite{CDMD1, CDMD2}. In the compression process, we started with a low-density hard-ellipse fluid at $\phi=0.005$ and used a quite small growth rate of $10^{-4}$ so that the created initial configuration already remained close to be in an equilibrium or stable state. These initial states were then used as inputs for EDMD simulations. At each aspect ratio, $10^{3}$ time units at $\phi\leq0.6$ and up to $10^{4}$ time units at $\phi>0.6$ were used to equilibrate the system, and then another $2\times10^{4}$ time units for collecting data. Four independent runs were performed for each state point in order to obtain reliable results and improve the statistics.Thus, the results presented in this paper have been averaged among four independent samples.. It is worth mentioning that the slowest run takes over four weeks on a $2.6$ GHz central processing unit (CPU).

We note that in the solid phase (i.e., particles show long-range order both positionally and orientationally), it will be useful to allow the shape of the simulation box to change during the simulation, because the box shape is necessarily compatible with the equilibrium shape of the unit cell in order to form a solid phase. This has been done in previous $NPT$ MC simulations~\cite{Frenkel2} in order to study the transition from an isotropic or a nematic phase to a solid phase in hard ellipses. The box shape is fixed to be square in our simulations since our aim here is not to determine the location where the solid phase forms, but to focus on the physical properties of the formed phases in EDMD simulations. Meanwhile, we do not investigate the melting process of a perfect hard-ellipse solid (which is a stretched triangular lattice). For these reasons, we did not observe the solid phase for all the state points studied in this work, please see Subsection III B for further discussion.

\section{Results and discussion}

In this section, we present and discuss our results of the thermodynamic, structural and self-diffusive properties in hard ellipses. First of all, the EOS is shown for a wide range of aspect ratios and the reduced pressure (i.e., compressibility factor) in the isotropic phase is compared with a recent result of the SPT. We also explore the dependence of the reduced pressure on the aspect ratio in the different phases and discuss the influence of the particle motions on the thermodynamics. In Subsection III B, we analyze in detail the positional and orientational order of hard ellipses and compare our results with previous simulations when possible. Finally, we investigate the self-diffusion of hard ellipses and demonstrate connections between the structure and the self-diffusion. The relevance of our results to the glass transition in hard ellipses is briefly discussed. We also show the results of the isodiffusivity lines in both the translational and the rotational degrees of freedom and discuss the effect of spatial dimension on the diffusive dynamics of hard ellipsoidal particles.

\subsection{Equation of state}

\begin{figure}[b]
 \centering
 \includegraphics[angle=0,width=0.45\textwidth]{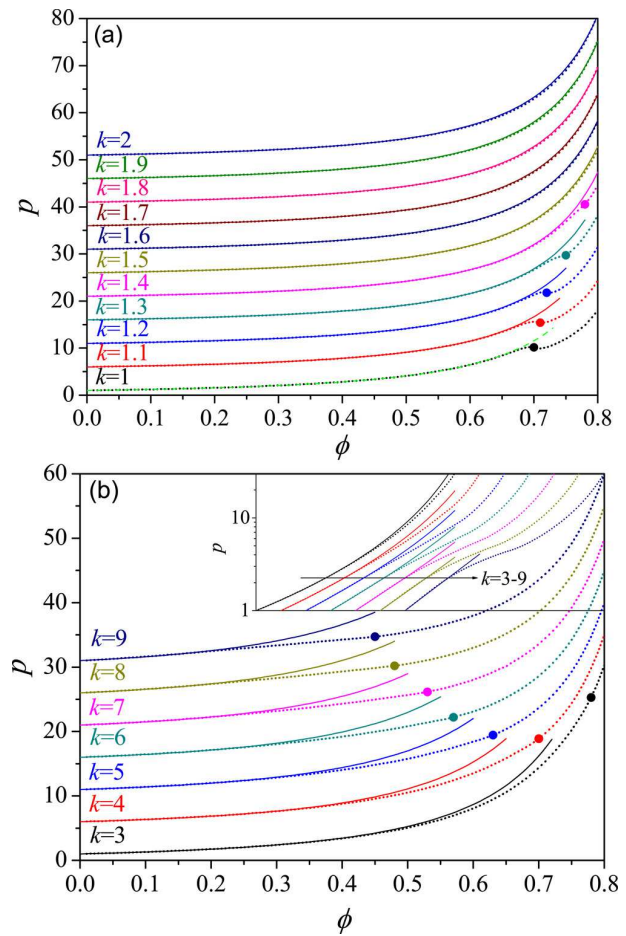}
 \caption{Reduced pressure $p$ as a function of area fraction $\phi$ for (a) hard disks and hard ellipses with $k=1.1-2$ and (b) hard ellipses with $k=3-9$. The results are shifted up by $5$ from the preceding small $k$ for clarity. The dotted lines are simulation results and the solid lines are the results of the SPT [Eq. (5)]. The green dashed line in (a) is the SPT prediction for the equation of state of hard disks in the isotropic state, i.e., Eq. (10) in Ref.~\cite{Boulik}. The circles indicate the area fractions where the system starts to form stable plastic crystals in (a) and stable nematic crystals in (b), respectively (see the structural analysis in Subsection III B). The semi-log plot in the inset of Fig. 1(b) highlights the change of dependence of $p$ on $\phi$ with increasing $k$.}
\end{figure}

\begin{figure}[tb]
 \centering
 \includegraphics[angle=0,width=0.45\textwidth]{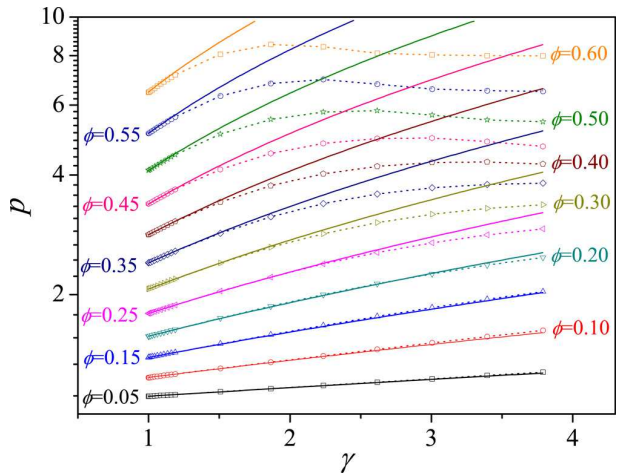}
 \caption{Reduced pressure $p$ as a function of $\gamma$ at fixed densities. The solid lines are the results of Eq. (5).}
\end{figure}

\begin{figure}[b]
 \centering
 \includegraphics[angle=0,width=0.45\textwidth]{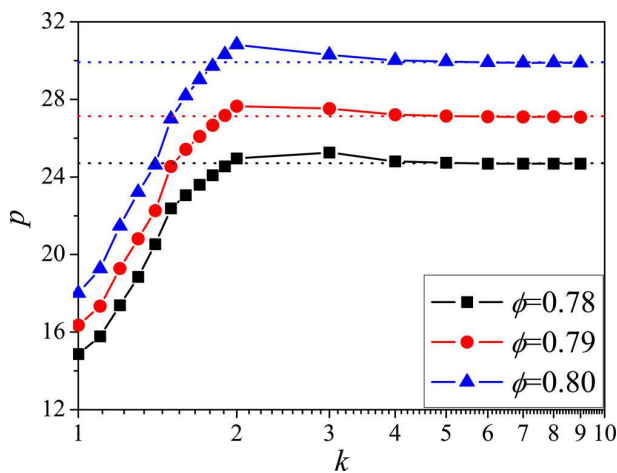}
 \caption{$k$-dependence of $p$ at the highest densities studied. The dotted lines are a guide to the eye. For these densities, the system has a plastic phase for $k\leq1.4$, a nematic phase for $k\geq3$ and an isotropic phase in between.}
\end{figure}

Knowledge of an accurate EOS for a model system is of fundamental importance and particularly useful for providing information on what type of an underlying phase transition is. We present the results of EOS for hard ellipses for various aspect ratios in Fig. 1. The EOS for a system of hard disks is also included in Fig. 1(a). Here, the reduced pressure $p=PV/Nk_{B}T$ is computed by the momentum exchange from particle collisions in EDMD (see Appendix A). We checked that our results for $k=2$, $4$ and $6$ are consistent with the compression runs in previous MC simulations~\cite{VieillardBaron, Frenkel2}.

Let us first focus on hard ellipses with small $k$ values. In this situation, the orientational order is not appreciable in the system but the positional order can develop upon compression, leading to the formation of plastic crystals at sufficiently high densities. The appearance of the positional order can significantly affect the thermodynamics of the system and it is indeed reflected by a discontinuity in the EOS, as shown in Fig. 1(a). The plastic crystal was already observed for a system of hard ellipses with $k=1.01$ about four decades ago by Vieillard-Baron~\cite{VieillardBaron}. Our results demonstrate that the plastic crystal can be formed in hard ellipses with aspect ratios up to $k=1.4$. The observation of the discontinuity in the EOS is compatible with the scenario of a first-order transition, but it should be viewed with some caution since the nature of the formation of a 2D crystalline phase is unclear even in the case of hard disks; e.g., although the Kosterlitz, Thouless, Halperin, Nelson and Young (KTHNY) theory may not rule out other melting scenarios~\cite{Nelson}, a famous one is that the melting of hard disks proceeds via two continuous phase transitions with an intermediate phase called hexatic phase~\cite{KT, HN, Young}, while the recent large-scale computer simulations~\cite{Bernard} reveal that the liquid-hexatic transition is of first order. We will further discuss this issue in Subsection III B by analyzing the structure of the system. Moreover, we find in Fig. 1(a) that the discontinuity shown in the EOS becomes less apparent as $k$ gets larger and nearly undetectable within the studied density range for $k\geq1.5$. This can be easily understood since the particle's anisotropy tends to destroy the positional order of the system, and thus will be in favor of the isotropic phase. For hard ellipses with sufficiently large elongations, the orientational order is expected to appear and the positional order will be highly suppressed, resulting in the formation of 2D nematic liquid crystals in hard ellipses. As already found in other 2D models~\cite{Frenkel1, Bates}, the thermodynamic properties of the system are indeed not very sensitive to the occurrence of the orientational order. This is also confirmed by our results of the EOS for hard ellipses with $k\geq3$, as shown in Fig. 1(b). For these $k$ values, the system does have a nematic phase above some area fraction, but the corresponding change in the EOS is rather mild. Thus, we cannot unambiguously determine the nature of the I-N transition based on solely the thermodynamics. However, by a close inspection of the results in the semi-log plot [see the inset of Fig. 1(b)], some difference in the EOS can be seen with varying $k$. It seems that some inflection points exhibit in the EOS for large aspect ratios, while the slope $d\log p/d\phi$ monotonically increases with $\phi$ for small aspect ratios, i.e., the $\phi$-evolution of $p$ differs qualitatively from relatively small to very large aspect ratio. This suggests that the nature of the I-N transition changes with aspect ratio. In fact, previous MC simulations imply the existence of a tricritical point between $k=4$ and $6$ in the phase diagram of hard ellipses~\cite{Frenkel2}. The I-N transition is first-order below the tricritical point and becomes continuous above it. Interestingly, we find that the appearance of the inflection points in the EOS starts within the same $k$ range. Thus, our results appear to support the existence of a tricritical point in the phase diagram of hard ellipses. However, since the location of such a point is very hard to determine even if it exists, we will focus on characterizing the properties of different phases in hard ellipses. Next, we compare the reduced pressure in the isotropic phase of hard ellipses with a recent result of the SPT.

As mentioned in Section I, the EOS of hard ellipses can also be obtained by various theories. Hence, it is meaningful to test these theoretical predictions as our results cover a wide window of $k$ and $\phi$ and contain data in the various phases. SPT in general provides a simple form for the EOS of a model system and its usefulness has been confirmed by the study of hard disks~\cite{SPT1, SPT2}. Other existing theories also provide relevant results, but the solution of the EOS needs to be solved numerically~\cite{Cuesta1, Ferreira, Varga, Schlacken}. Thus, we concentrate here on a very recent result of the SPT~\cite{Boulik}. On the basis of the SPT, Boul\'{i}k derived an EOS for the isotropic fluid of hard ellipses~\cite{Boulik}, which has the following form
\begin{equation}
p=\frac{PV}{Nk_{B}T}=\frac{1}{(1-\phi)}+\frac{\gamma\phi[1+\gamma(\phi/7-\phi^{2}/14)]}{(1-\phi)^{2}},
\end{equation}
where the non-circularity parameter $\gamma$ has been introduced. $\gamma$ is calculated according to the formula $\gamma=C^{2}/(4\pi^{2}ab)$ with $C$ the perimeter of the ellipse. The value of $C$ can be expressed accurately by the complete elliptic integral of the second kind, and a rather good approximation can be obtained from the following formula:
\begin{equation}
C\approx\pi(a+b)(1+\frac{3\nu^{2}}{10+\sqrt{4-3\nu^{2}}}),
\end{equation}
where $\nu=(a-b)/(a+b)$. We use Eq. (6) to calculate $\gamma$ in this work and the results of the Eq. (5) are shown as solid lines in Fig. 1. As a reference, the area fractions where the system starts to form stable plastic crystals and stable nematic crystals are also shown in Fig. 1. As can be seen in Fig. 1(a), the simulation results and the SPT prediction are almost indistinguishable in the isotropic phase for $k\leq2$, indicating a fairly good description of Eq. (5) for the EOS of hard ellipses with small $k$ values. Meanwhile, the predicted EOS deviates from the simulation results above some area fraction for $k\geq3$, as shown in Fig. 1(b). As explained in Ref.~\cite{Boulik}, the system is not isotropic but exhibits orientational order at high densities for large $k$ values. In Fig. 1(b), one can find that the area fraction where Eq. (5) starts to deviate from the simulations is much smaller than the I-N transition point [see circles in the main plot of Fig. 1(b)]. However, it should be emphasized that the short-range orientational order already appears in the system at smaller area fractions (see Subsection III B). Taking this into account, we find that Eq. (5) well describes the isotropic branch of the EOS for hard ellipses even with large $k$ values. We can also use $\gamma$ as a variable to test Eq. (5). The results for representative area fractions are shown in Fig. 2. We find that the theoretical and simulation results agree well for low densities. At high densities, deviations appear for large $\gamma$ values because of the occurrence of short-range or some long-range orientational order, as evidenced by the non-monotonic dependence of $p$ on $\gamma$. Therefore, we conclude that SPT provides a good description of the EOS for the isotropic phase of hard ellipses.

It is interesting to further explore dependence of the reduced pressure on the aspect ratio in the various phases of hard ellipses. To this end, we plot $p$ as a function of $k$ for $\phi=0.78$, $0.79$ and $0.8$ in Fig. 3. For these densities, the system has a plastic phase for $k\leq1.4$, a nematic phase for $k\geq3$ and an isotropic phase in between (see Subsection III B). We find that $p$ monotonically increases with $k$ at fixed densities in both the isotropic and the plastic phases (i.e., $k\leq2$ at $\phi\geq0.78$), suggesting that the thermodynamics of the system can be significantly influenced by the particle shape when the ellipses are oriented at random. Note that this conclusion is still valid for even larger $k$ values (see the results for $\phi<0.2$ in Fig. 2). Interestingly enough, in the nematic phase (i.e., $k\geq3$ at $\phi\geq0.78$), when the particles show some long-range orientational order but have their centers of mass at random, $p$ depends on $k$ very weakly. Although not pointed out before, the same conclusion can be drawn from previous $NPT$ MC simulations~\cite{Frenkel2}, which indicate that the area fractions are almost identical at the same pressure in the nematic phase of hard ellipse at least for $k=4$ and $6$. To understand the above findings, we recall that the reduced pressure is calculated from the momentum exchange during collisions of two ellipses, which indeed contains translational and rotational parts (although they are coupled with each other). It is expected that the contribution from the rotational motions becomes more apparent as the ellipses get more elongated. This is the reason why $p$ increases with $k$ in the plastic and the isotropic phases, where the particles can freely move both translationally and rotationally (see Fig. 14). In the nematic phase, however, particles on average line up preferentially along a common direction. In this case, the rotational dynamics becomes extremely slow, but the motions of the particles have little hindrance in the translational degrees of freedom (see Fig. 14) and thus the translational part becomes dominant in determining the reduced pressure of the system. We then speculate that the translational part must be very similar for all aspect ratios at the same density in the nematic phase so that $p$ has only weak dependence on $k$. Therefore, the thermodynamics of hard ellipses could be explained in terms of the particle motions.

\subsection{Static structure}

We now turn to the structural properties of hard ellipses. As a system of hard ellipses with sufficiently large aspect ratios has a transition from the isotropic liquid to the 2D liquid crystal, much attention has been paid on the structure of the nematic phase in hard ellipses, i.e., on characterizing the orientational order of elongated hard ellipses. Instead of focusing solely on the orientational correlations, we explore both positional and orientational order in the various phases of hard ellipses in this subsection.

\begin{figure}[tb]
 \centering
 \includegraphics[angle=0,width=0.45\textwidth]{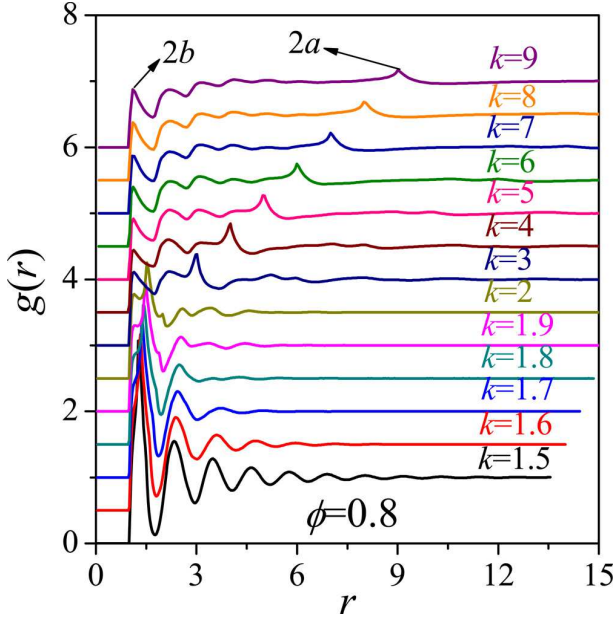}
 \caption{Pair correlation function $g(r)$ of hard ellipses at $\phi=0.8$ for $k\geq1.5$. The results are shifted up by $0.5$ from the preceding small $k$ for clarity. As there are two basic length scales in a system of hard ellipses, $g(r)$ also peaks at $r=2a$, as indicated by the arrows for $k=9$.}
\end{figure}

\begin{figure}[b]
 \centering
 \includegraphics[angle=0,width=0.45\textwidth]{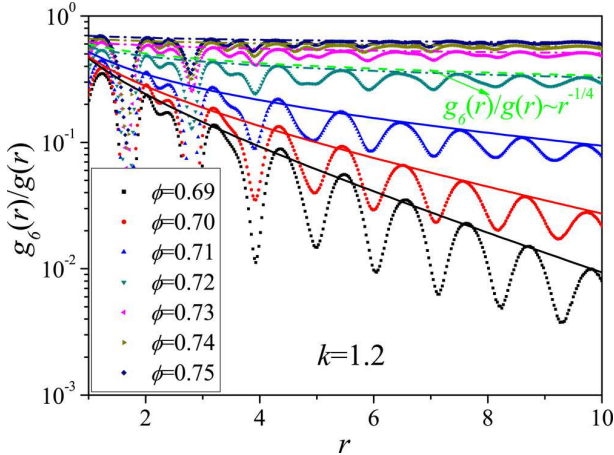}
 \caption{$g_{6}(r)/g(r)$ of hard ellipses with $k=1.2$ in the vicinity of the isotropic-plastic transition. The solid lines are the results of the OZ fittings and the dash dotted lines are the results of the power-law fittings. The green dashed line indicates $g_{6}(r)/g(r)\sim r^{-1/4}$. The results are similar for other aspect ratios with $k\leq1.4$.}
\end{figure}

\begin{figure}[t]
 \centering
 \includegraphics[angle=0,width=0.45\textwidth]{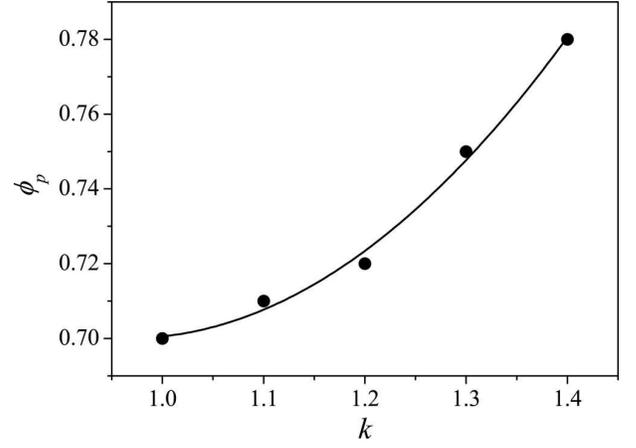}
 \caption{Isotropic-plastic transition point $\phi_{p}$ as a function of $k$. The solid line is a guide to the eye.}
\end{figure}

\begin{figure}[b]
 \centering
 \includegraphics[angle=0,width=0.45\textwidth]{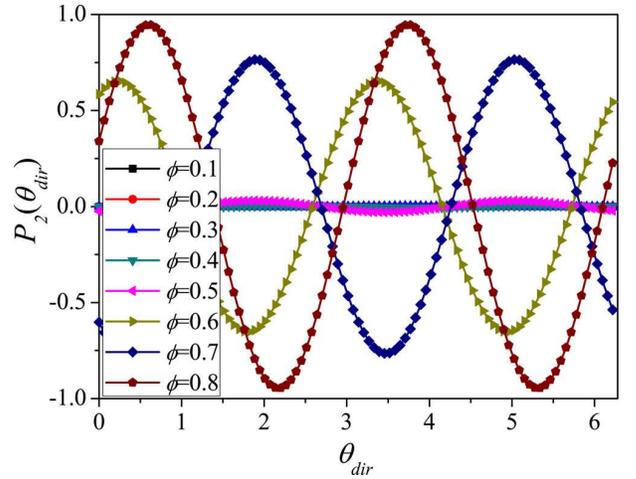}
 \caption{Nematic order parameter $P_{2}$ as a function of nematic director $\theta_{dir}$ for hard ellipses with $k=6$ for various $\phi$. Note that the result is shown for a single sample because $P_{2}(\theta_{dir})$ cannot be averaged among different samples since the nematic direction differs from one sample to another.}
\end{figure}

\begin{figure}[t]
 \centering
 \includegraphics[angle=0,width=0.45\textwidth]{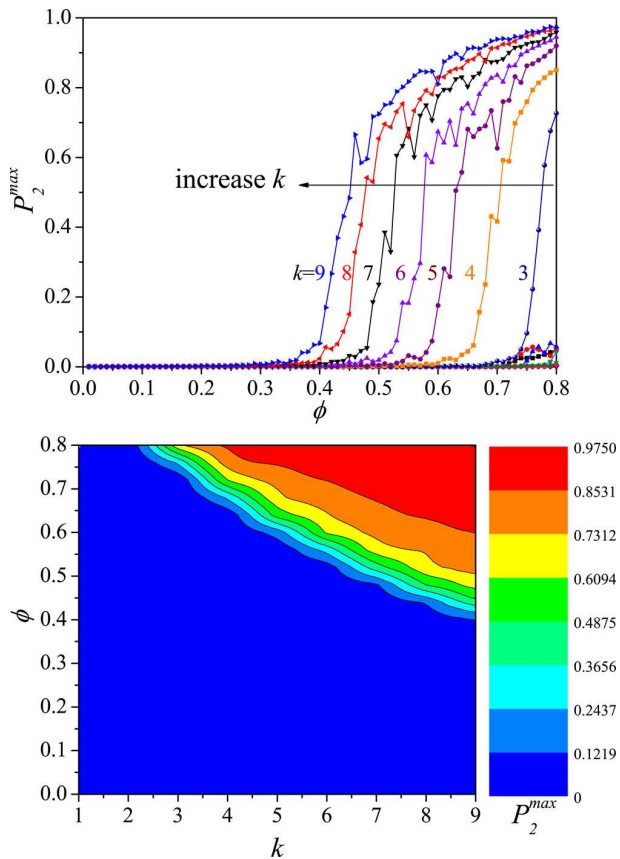}
 \caption{Upper: $P_{2}^{max}$ as a function of $\phi$ for various $k$. Lower: contour plot of $P_{2}^{max}$ in the plane of $\phi$ and $k$.}
\end{figure}

\begin{figure}[b]
 \centering
 \includegraphics[angle=0,width=0.45\textwidth]{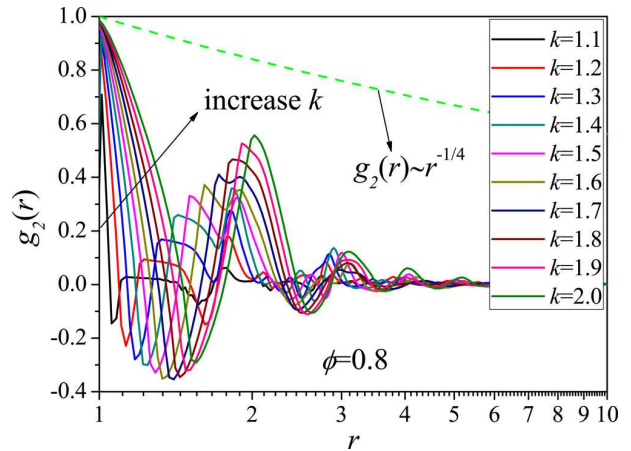}
 \caption{Angular correlation function $g_{2}(r)$ for hard ellipses with $k\leq2$ at $\phi=0.8$. The green dashed line indicates $g_{2}(r)\sim r^{-1/4}$.}
\end{figure}

\begin{figure}[t]
 \centering
 \includegraphics[angle=0,width=0.45\textwidth]{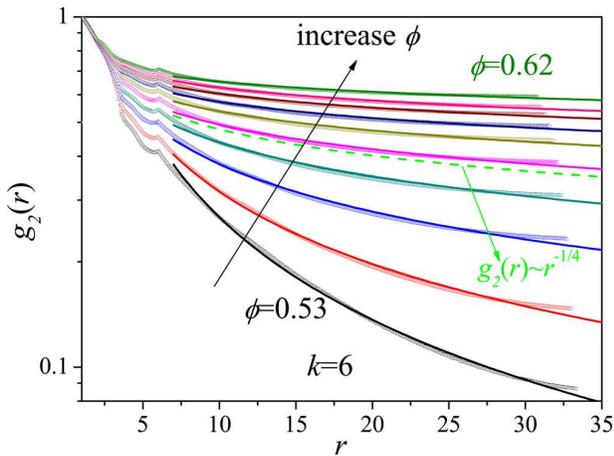}
 \caption{Angular correlation function $g_{2}(r)$ for a system of hard ellipses with $k=6$ in the vicinity of the isotropic-nematic transition. The solid lines are the results of the power-law fittings. The green dashed line indicates $g_{2}(r)\sim r^{-1/4}$. The results are similar for other aspect ratios with $k\geq3$. }
\end{figure}

\begin{figure}[b]
 \centering
 \includegraphics[angle=0,width=0.45\textwidth]{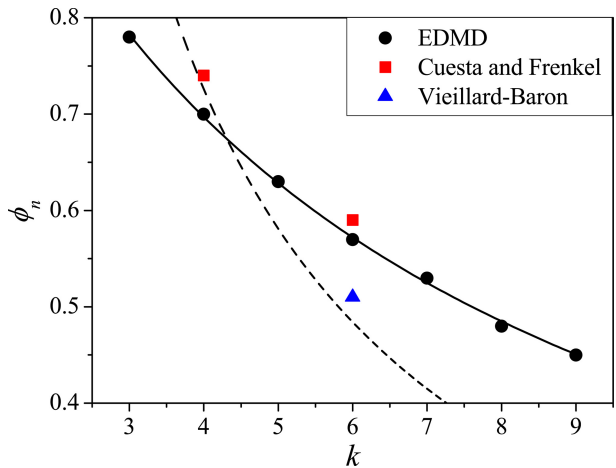}
 \caption{Isotropic-nematic transition point $\phi_{n}$ as a function of $k$. The triangle and the squares are the MC results taken from Ref.~\cite{VieillardBaron} and Ref.~\cite{Frenkel2}, respectively. The dashed and solid lines are the fits to the EDMD data by $\phi_{n}=A/k$ with $A=2.90$ and $\phi_{n}=\phi_0/(k_0+k)$ with $\phi_0=6.37$ and $k_0=5.14$, respectively.}
\end{figure}

\begin{figure}[t]
 \centering
 \includegraphics[angle=0,width=0.45\textwidth]{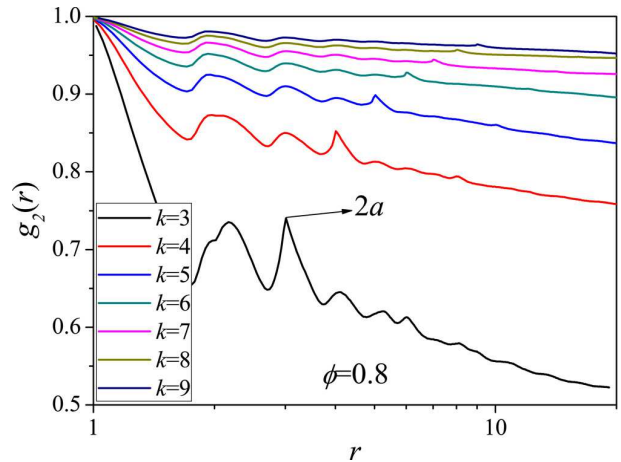}
 \caption{Angular correlation function $g_{2}(r)$ for hard ellipses with $k\geq3$ at $\phi=0.8$.}
\end{figure}

\begin{figure}[t]
 \centering
 \includegraphics[angle=0,width=0.45\textwidth]{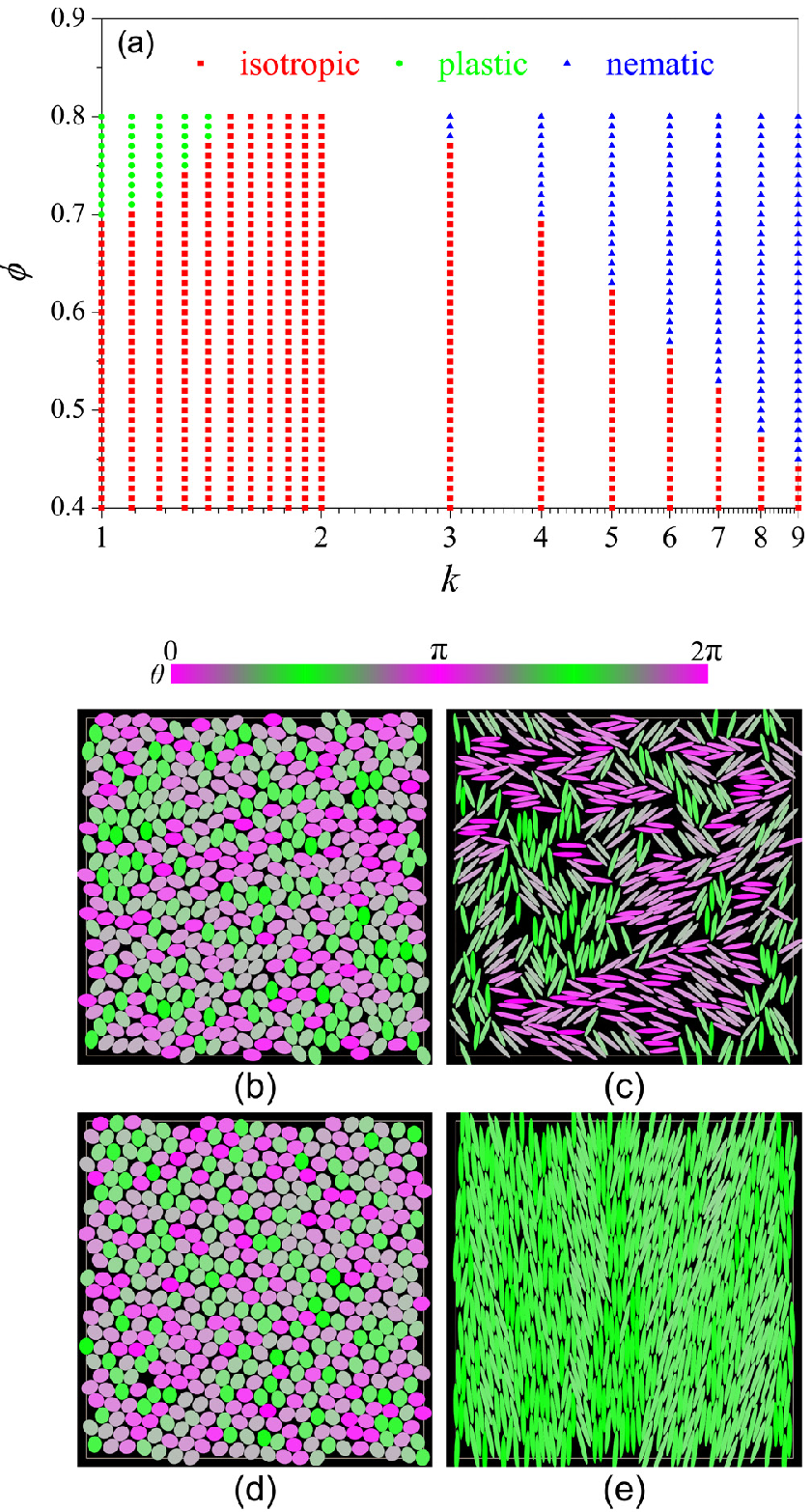}
 \caption{(a) Phase diagram of hard ellipses in the plane of aspect ratio $k$ and area fraction $\phi$. (b)-(e) Representative snapshots for a high-density isotropic phase with $k=1.5$ and $\phi=0.8$, a low-density isotropic phase with $k=6$ and $\phi=0.5$,  a plastic phase with $k=1.2$ and $\phi=0.8$, and a nematic phase with $k=9$ and $\phi=0.8$. Note that there is visually no distinction for the orientation of an ellipse with $\theta$ and $\theta+\pi$ so that particles with $\theta$ are shown as the same color as those with $\theta+\pi$ in the snapshots.}
\end{figure}

We first consider the pair correlation function of hard ellipses, which is defined as
\begin{equation}
g(r)=\frac{L^2}{2\pi r\Delta rN(N-1)}<\sum_{j \neq k}\delta(r-|\textbf{r}_{jk}|)>,
\end{equation}
where $r$ is the distance between the centers of mass of the particles. In general, characteristic peaks will appear in $g(r)$ at large distances if the system possesses some long-range positional order. This is indeed observed in systems of hard ellipses with $k\leq1.4$ at high densities (data not shown), where the centers of mass of the particles form a triangular (or hexagonal) lattice.  In hard ellipses with larger aspect ratios, an ordered structure in principle can also form translationally although it does not need to be a perfect triangular one. We present the results of $g(r)$ at $\phi=0.8$ for $k\geq1.5$ in Fig. 4. Note that there are two basic length scales (i.e., the semi-major axis $a$ and the semi-minor axis $b$) in a system of hard ellipses so that $g(r)$ also peaks at the distance of $2a$. Beyond the distance of $2a$, we find from Fig. 4 that $g(r)$ rapidly decays to the value of $1$, indicating the absence of any long-range translational order even at the largest density studied. The reason, why we did not observe the formation of a translationally ordered structure for large aspect ratios, has been explained in Section II. Moreover, it is clear that several peaks are also shown in $g(r)$ at between $2b$ and $2a$, suggesting the existence of some short-range translational order. As we shall see later, this is indeed due to the appearance of the orientational order.

The pair correlation function only provides a qualitative way to see whether a system possesses long-range positional order or not. Meanwhile, we can study the bond-orientation correlation functions~\cite{Comment}, in order to quantitatively identify the location of the transition from the isotropic liquid to the hexagonal crystal. We first define the sixfold bond-orientation order parameter
\begin{equation}
\psi_{6}^{j}=\frac{1}{n_{j}}\sum_{m=1}^{n_{j}}\exp(i6\theta_{m}^{j}),
\end{equation}
where $i=\sqrt{-1}$, $n_{j}$ is the number of the neighbors of the $j$th particle and $\theta_{m}^{j}$ is the angle between the vector $(\textbf{r}_{m}-\textbf{r}_{j})$ and the $x$ axis. Here, we define two ellipses to be neighbors if they overlap when uniformly expanding their sizes by $1.4$. Then, the spatial correlation of $\psi_{6}^{j}$ can be calculated by
\begin{equation}
g_{6}(r)=\frac{L^2}{2\pi r\Delta rN(N-1)}<\sum_{j\neq k}\delta(r-|\textbf{r}_{jk}|)\psi_{6}^{j}\psi_{6}^{k*}>.
\end{equation}
The quasi-long-range bond-orientation order is evidenced by an algebraic decay of $g_{6}(r)/g(r)$, i.e.,
\begin{equation}
g_{6}(r)/g(r)\sim r^{-\eta_{6}}.
\end{equation}
In particular, according to the KTHNY theory~\cite{HN}, $\eta_{6}$ has a value of $1/4$ at the boundary between the liquid phase and the hexatic phase. We thus use the criterion $g_{6}(r)/g(r)\sim r^{-1/4}$ to separate the plastic states from the isotropic states for hard ellipses with $k\leq1.4$. In the liquid phase, the system only exhibits short-range bond-orientation order and $g_{6}(r)/g(r)$ should decay exponentially. However, for high densities and especially in vicinity of the formation of the hexatic phase, we find that the bond-orientation correlation function can indeed be better described by the Ornstein-Zernike (OZ) function
\begin{equation}
g_{6}(r)/g(r)\sim r^{-1/2}\exp(-r/\xi_{6}),
\end{equation}
with $\xi_{6}$ the static correlation length. In fact, it is observed that the OZ function can well characterize the behavior of $g_{6}(r)/g(r)$ in 2D supercooled liquids~\cite{Tanaka1, Tanaka2, Xu1, Xu3}. We confirm the above claims by presenting the results of $g_{6}(r)/g(r)$ for a system of hard ellipses with $k=1.2$ in Fig. 5. As can be seen, $g_{6}(r)/g(r)$ decays slowly with increasing $\phi$. Below $\phi \approx 0.72$, the envelopes of $g_{6}(r)/g(r)$ can be well fitted by the OZ function. With further increasing $\phi$, a power-law decay with an exponent of $\eta_{6}<1/4$ holds for the behavior of $g_{6}(r)/g(r)$, indicating the emergence of quasi-long-range bond-orientation order. We thus estimate the location of the isotropic-plastic transition to be $\phi_{p}=0.72$ for $k=1.2$. Note that $\phi_{p}$ obtained in this way is actually an upper bound to the exact value. The values of $\phi_{p}$ for other aspect ratios are shown in Fig. 6. and we confirmed that the obtained value for hard disks (i.e., $k=1$) is in agreement with other studies within the simulation accuracy~\cite{Bernard}. The above analysis indicates that hard ellipses for $k\leq1.4$ share the similar behavior with hard disks on the structure as well as the thermodynamics (see discussion in Subsection III A). Thus, the nature of the isotropic-plastic transition of hard ellipses should be the same as that of hard disks.

Turning to the orientational order of hard ellipses, we first focus on the nematic order parameter as a function of $\phi$ and $k$. The 2D nematic order parameter is defined as
\begin{equation}
P_{2}=\frac{1}{N}<\sum_{j=1}^{N}2\cos^{2}(\theta_{j}-\theta_{dir})>-1,
\end{equation}
where $\theta_{j}$ is the angle characterizing the orientation of the $j$th ellipse with respect to the $x$ axis and $\theta_{dir}$ the orientation of the nematic director. Note that this definition is equivalent to Eq. (3). As the nematic director is not known \emph{a priori}, $P_{2}$ and $\theta_{dir}$ are usually determined by finding the eigenvalues and the eigenvectors of a tensor order parameter~\cite{Frenkel1, Frenkel2}. As pointed out and observed in previous work~\cite{Frenkel1, Frenkel2}, the nematic order parameter obtained from the above method has a clearly nonzero value even in the isotropic phase due to the finite system size used in simulations. Instead of using the tensor order parameter, we set $\theta_{dir}$ as a variable and calculate $P_{2}$ as a function of $\theta_{dir}$ in this work. We show the results of $P_{2}(\theta_{dir})$ for various area fractions for $k=6$ in Fig. 7 as an illustration. As can be seen, $P_{2}(\theta_{dir})$ shows extrema, and the absolute values of these extrema are the same (we denote it by $P_{2}^{max}$) because of the periodicity of the cosine function in the definition of $P_{2}$. We then use $P_{2}^{max}$ to quantify the nematic order of hard ellipses. In Fig. 8, $P_{2}^{max}$ is presented as a function of $\phi$ for various aspect ratios and also shown as a contour plot in the plane of $\phi$ and $k$. We find from Fig. 8 that $P_{2}^{max}$ is nearly zero at low densities and has clearly nonzero values at high densities for $k\geq3$. Thus, it appears that $P_{2}^{max}$ is more sensitive to the onset of the orientational order than $q$. For $k\leq2$, a clearly nonzero but still very small value of $P_{2}^{max}$ sets in above $\phi\approx0.7$, implying the absence of any long-range orientational order. The quasi-long-range orientational order is expected to appear for systems of hard ellipses with larger aspect ratios, as confirmed by the results of $P_{2}^{max}$ for $k\geq3$. Moreover, the area fraction where the system starts to orientate becomes small as $k$ increases, revealing the fact that a system of more elongated hard ellipses has a stronger tendency to form nematic liquid crystals.

It is interesting to explore whether $P_{2}^{max}$ is dependent on the system size in the nematic phase. Unlike the order parameter $q$ computed from the tensor order parameter (dependence of the system size on $q$ has been confirmed by MC studies~\cite{Frenkel1, Bates, Frenkel2}), we find that the system size doesn't strongly influence the value of $P_{2}^{max}$ for both the isotropic and the nematic phases studied (see Appendix B). Hence, it is not proper to estimate the point of the I-N transition from the dependence of the system size on $P_{2}^{max}$. We can quantitatively determine the location of the I-N transition $\phi_{n}$ from the angular correlation function defined by Eq. (4). We mainly concentrate on the case of $l=1$, i.e.,
\begin{equation}
g_{2}(r)=<\cos(2[\theta(r)-\theta(0)])>\sim r^{-\eta_{2}},
\end{equation}
where $\eta_{2}=2k_{B}T/\pi K$ and the average is performed over all pairs with the distance $r$. As introduced in Section I, the stable nematic phase is then identified by a power-law decay with $\eta_{2}<1/4$ for $g_{2}(r)$. The results of $g_{2}(r)$ are shown for $k\leq2$ at $\phi=0.8$ in Fig. 9. Consistent with the results of $P_{2}^{max}$, we find that $g_{2}(r)$ decays much faster than $g_{2}(r)\sim r^{-1/4}$ even at the largest studied density for these aspect ratios, indicating again there is no quasi-long-range orientational order in the system. The algebraic decay appears at high densities for $k\geq3$, as evidenced in Fig. 10 for a system of hard ellipses with $k=6$. We then use the power law $g_{2}(r)\sim r^{-\eta_{2}}$ to fit $g_{2}(r)$ at $r>2a$ and estimate the location of the I-N transition $\phi_{n}$ to be the area fraction when the value of $\eta_{2}$ is smaller than $1/4$. Again, $\phi_{n}$ obtained in this way is actually an upper bound to the exact value. It should also be stressed that the I-N transition in a finite system tends to occur at a lower density than in an infinite system because the Frank's constant obtained from simulations on small systems is larger than that in the infinite system size~\cite{Frenkel1}. We show the results of $\phi_{n}$ as a function of $k$ in Fig. 11. As can be expected, $\phi_{n}$ decreases with increasing $k$. Comparing with previous MC results, we find that $\phi_{n}$ for $k=6$ obtained in this work is larger than that in Ref.~\cite{VieillardBaron} but close to the result in Ref.~\cite{Frenkel2}. The reason, why the I-N transition was observed at a smaller area fraction by Vieillard-Baron, has been hinted by Frenkel and Eppenga~\cite{Frenkel1}. Similar to the case of $k=6$, the value of $\phi_{n}$ for $k=4$ in this work is also a little smaller than that of Ref.~\cite{Frenkel2}. The slight difference between the results of Ref.~\cite{Frenkel2} and ours is possibly due to the different particle number and different method used ($N\approx200$ and $NPT$ MC in Ref.~\cite{Frenkel2} versus $N=500$ and $NVT$ MD in this work). We also note a recent MC study~\cite{Davatolhagh}, where $\phi_{n}$ for $k=9$ is roughly estimated to be $0.58$, which is much larger than ours.

Our results also provide an opportunity to test theoretical predictions on the $k$-dependence of $\phi_{n}$ for hard ellipses. Several theories~\cite{Cuesta1, Cuesta2, Ferreira, Varga, Schoot, Schlacken} have predicted the occurrence of the I-N transition in hard ellipses. In particular, the Onsager theory-based approach~\cite{Schoot} and SPT~\cite{Schlacken} predict that the relation $\phi_{n}\sim 1/k$ emerges for sufficiently elongated hard ellipses. Since the values of $k$ studied in this work are not large enough and the hard-needle regime has probably not been reached yet, such a relation cannot well describe our results (see the dashed line in Fig. 11). Instead, we find that a different form, $\phi_{n}=\phi_0/(k_0+k)$ with $\phi_0$ and $k_0$ adjustable parameters, provides a fairly good description of our data, as shown by a solid line in Fig. 11 with $\phi_0=6.37$ and $k_0=5.14$. 

As mentioned in Section I, the nature of the orientational order in the 2D nematic phase is also a central theme in the study of liquid crystals. The algebraic decay of $g_{2}(r)$ already indicates the lack of the true long-range orientational order in the nematic phase of hard ellipses. We further confirm that the algebraic orientational order holds even for the largest studied density, as shown in Fig. 12, although $g_{2}(r)$ may decay very slowly in this case. Thus, our simulations strongly support that the nematic phase in hard ellipses has quasi-LRO.

Based on the above structural analysis, we present a phase diagram of hard ellipses and several representative snapshots in Fig. 13. Our simulations show three different phases within the investigated state points. For $k\leq1.4$, the system has a transition from a low-density isotropic liquid to a plastic crystal at high densities [Fig. 13(d)]. With increasing the aspect ratio up to $k=2$, we did not observe any phase transition within the whole density range studied in this work [Fig. 13(b)]. For hard ellipses with $k\geq3$, an isotropic phase [Fig. 13(c)] will transform into a nematic phase [Fig. 13(e)] at sufficiently high densities. Again, we confirm from the snapshot in Fig. 13(c) that some short-range orientational order exists even in the isotropic phase of elongated hard ellipses. On the other hand, although the simulated phase diagram does not include any solid phase, we should stress again that hard ellipses do have the transition from an isotropic or a nematic phase to a solid phase, as revealed in previous MC simulations~\cite{VieillardBaron, Frenkel2}. We even expect that a transition from a plastic crystal to a solid phase will occur for small $k$ values at even larger densities than $\phi=0.8$. The reason why such a transition was not observed in our work, as explained in Section II, is due to the fact that a solid is hard to spontaneously form from a liquid in $NVT$ MD simulations. On the other hand, it will be very interesting to explore the high-density part of the phase diagram for hard ellipses, where the stable crystalline phase will exist. In fact, our knowledge is very limited so far about the equilibrium structure of the solid in hard ellipsoidal particles, although it can be readily recognized that the densest packing of hard ellipses has the same density as the densest packing of hard disks and has a stretched triangular lattice structure~\cite{Donev2}. However, such explorations are hindered by the present method. More elegant methods will be needed to clarify this issue, which remains a challenge from both theoretical and computational points of view.

\subsection{Self-diffusion}

Self-diffusion is of great importance in many real processes, but it is still poorly understood in hard ellipses. The diffusive property of hard ellipses with $k=9$ has been studied recently by experiments~\cite{Zheng1} and MC simulations~\cite{Davatolhagh}. Yet, the effects of aspect ratio and area fraction on the self-diffusion of hard ellipses are unclear. In this subsection, we investigate in detail the self-diffusion of hard ellipses over a wide range of aspect ratios and area fractions, including the isotropic, plastic and nematic phases, and demonstrate the connections between the structure and the self-diffusion. We further discuss the relevance of our results to the glass transition in hard ellipses and provide the results of the isodiffusivity lines.

\begin{figure}[tb]
 \centering
 \includegraphics[angle=0,width=0.45\textwidth]{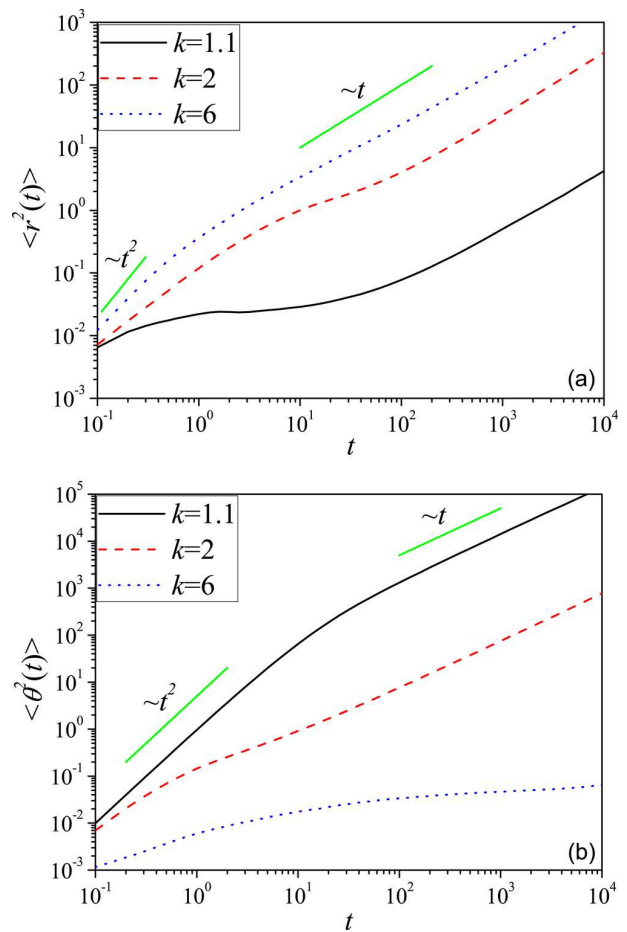}
 \caption{Time evolution of (a) translational and (b) rotational MSDs at $\phi=0.8$ for three aspect ratios. The system has a plastic phase for $k=1.1$, an isotropic phase for $k=2$ and a nematic phase for $k=6$.}
\end{figure}

The self-diffusion is measured here by the translational and rotational mean squared displacements (MSD)
\begin{equation}
<r^{2}(t)>=\frac{1}{N}<\sum_{j=1}^{N}|\textbf{r}_{j}(t)-\textbf{r}_{j}(0)|^{2}>,
\end{equation}
\begin{equation}
<\theta^{2}(t)>=\frac{1}{N}<\sum_{j=1}^{N}|\theta_{j}(t)-\theta_{j}(0)|^{2}>,
\end{equation}
and the corresponding diffusion constants obtained from the long-time data of the MSD
\begin{equation}
D_{T}=\lim_{t\rightarrow\infty}\frac{<r^{2}(t)>}{4t},
\end{equation}
\begin{equation}
D_{\theta}=\lim_{t\rightarrow\infty}\frac{<\theta^{2}(t)>}{2t}.
\end{equation}
We first focus on the behavior of the MSD in the different phases of hard ellipses. The results of $<r^{2}(t)>$ and $<\theta^{2}(t)>$ are shown at $\phi=0.8$ for the plastic ($k=1.1$), isotropic ($k=2$) and nematic ($k=6$) phases of hard ellipses in Fig. 14. As can be seen, both translational and rotational motions of particles are ballistic at short times, and cross over into diffusive behavior at sufficiently long times in both the plastic and the isotropic phases. Moreover, an apparent subdiffusive behavior is seen in the intermediate time regime of the translational MSD due to the cage effect in the plastic phase [see the solid line in Fig. 14(a)]. By contrast, the rotational diffusion of particles becomes extremely slow in the nematic phase, while the translational MSD can still retain a diffusive regime at long times within the simulation time window (see the dotted lines in Fig. 14). We have seen in Subsection A that the above features shown in the particle motions may be responsible for the thermodynamics in the different phases of hard ellipses.

\begin{figure}[tb]
 \centering
 \includegraphics[angle=0,width=0.45\textwidth]{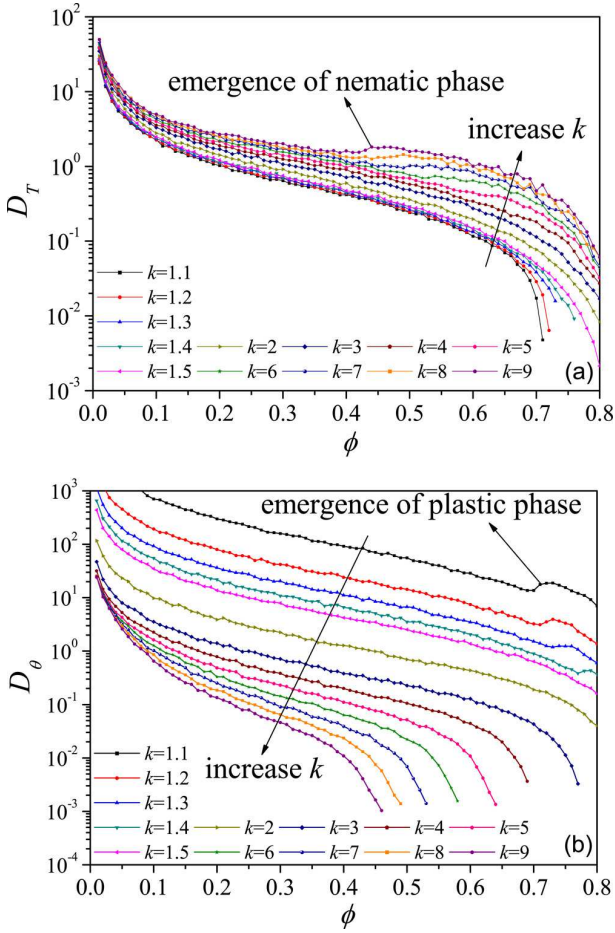}
 \caption{(a) Translational diffusion constant $D_{T}$ and (b) rotational diffusion constant $D_{\theta}$ as a function of area fraction $\phi$ for various $k$.}
\end{figure}

\begin{figure}[tb]
 \centering
 \includegraphics[angle=0,width=0.45\textwidth]{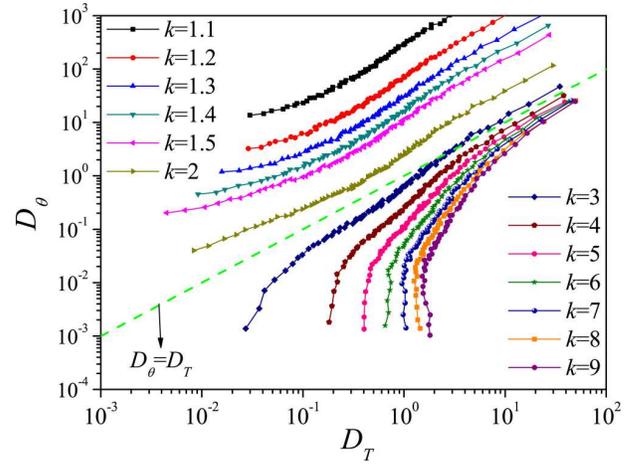}
 \caption{Rotational diffusion constant $D_{\theta}$ as a function of translational diffusion constant $D_{T}$ for various $k$. The green dashed line indicates $D_{\theta}=D_{T}$.}
\end{figure}

To better illustrate the effects of aspect ratio and area fraction on the self-diffusion of hard ellipses, we extract the translational and rotational diffusion constants from those of the corresponding MSDs retaining a diffusive regime. The results of $D_{T}$ and $D_{\theta}$ are shown in Fig. 15. At fixed area fractions, we observe that increasing $k$ leads to a slowing down of the rotational diffusion, while $D_{T}$ increases with aspect ratio. Moreover, the occurrence of the phase transition seems not to affect these trends at least within the studied density range. Then, a natural question is how self-diffusion responds to the phase transition. In a recent MD study~\cite{Davatolhagh}, it was found for a system of hard ellipses with $k=9$ that if the diffusion constants are normalized by the values in the infinite dilution limit, then both diffusion constants still monotonically decrease with $\phi$ but the normalized $D_{T}$ will exceed the normalized $D_{\theta}$ at the area fraction where the I-N transition takes place, i.e., the translational diffusion is enhanced with respect to the rotational diffusion as a result of the appearance of the quasi-long-range orientational order. We didn't observe crossing of the two normalized curves for the same aspect ratio by performing the same analysis (data not shown). Instead, our simulations reveal that the self-diffusion responds to the phase transition in a rather straight manner, which can be detected already in each curve; i.e., a sudden increase of $D_{\theta}$ is seen when a plastic crystal forms, as evidenced by a small peak in $D_{\theta}$ in Fig. 15(b), while the emergence of a nematic phase leads to an enhancement of $D_{T}$ although this observation seems to become less evident as $k$ gets smaller. The small peaks in Fig. 15 immediately indicate that the formation of the orientational or positional order will increase the translational or rotational mobilities of particles. Hence, a phase transition in the translational degrees of freedom is reflected by a corresponding change in the rotational diffusion and vice versa. We thus uncover a clear connection between the structure and the self-diffusion in hard ellipses. It is interesting to further explore the effect of the particle's anisotropy on the self-diffusion in each degrees of freedom by plotting $D_{\theta}$ as a function of $D_{T}$ at varying $k$. As shown in Fig. 16, $D_{\theta}$ is larger than $D_{T}$ within the full density range for $k\leq2$ and the particles diffuse slower in the rotational degrees of freedom than in the translational degrees of freedom for $k>3$. In addition, the curves show an upturn at high densities for $k\leq2$, while they turn down for $k\geq3$. This means that the translational mobility of hard ellipses will be similar with the rotational mobility at some aspect ratio between $2$ and $3$. If a glass is allowed to form in hard ellipses, we then expect that the rotational glass transition sets in at a lower density than the translational glass transition for $k\gtrsim3$, which was indeed observed recently in experiments of quasi-2D hard ellipsoids with $k\approx6$~\cite{Zheng}. Instead, the translational glass transition density will be smaller than the rotational one for $k<2$. This immediately suggests that there is a point in the phase diagram of glassy hard ellipses, where the translational and the rotational glass transition lines will intersect. Therefore, glasses in hard ellipses can be classified into three categories: a plastic glass in which dynamic arrest occurs in the translational degrees of freedom but the system behaviors as a liquid in the rotational degrees of freedom, a liquid glass in which the system becomes glass in the rotational degrees of freedom but the translational motions of particles remains ergodic, and a glass in which motions of the particles are dynamically arrested in both the translational and the rotational degrees of freedom.

\begin{figure}[tb]
 \centering
 \includegraphics[angle=0,width=0.45\textwidth]{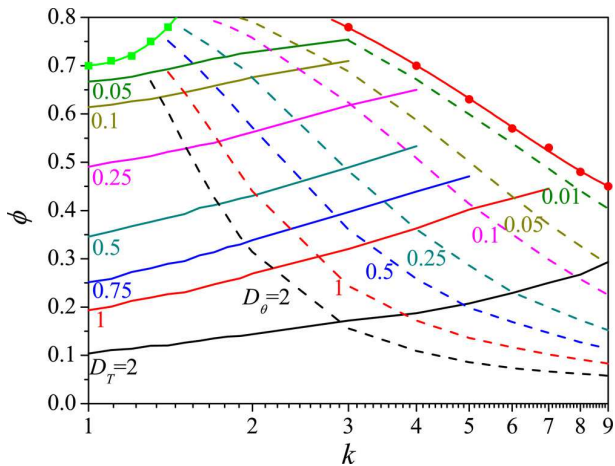}
 \caption{Isodiffusivity lines in the plane of aspect ratio $k$ and area fraction $\phi$. The solid lines are isodiffusivity lines from translational diffusion coefficients $D_{T}$ and the dashed lines are isodiffusivity lines from rotational diffusion coefficients $D_{\theta}$. The green squares and the red circles indicate the locations of the isotropic-plastic transition and the isotropic-nematic transition, respectively.}
\end{figure}

Another relevant property is the isodiffusivity of hard ellipses. As there are related results for hard ellipsoids of revolution~\cite{DeMichele}, we can evaluate such isodiffusivity lines for hard ellipses by proper interpolation and thus explore the qualitative effect of spatial dimension. We present the results of the isodiffusivity lines in Fig. 17. We find that an almost perpendicular crossing of the translational and the rotational isodiffusivity lines is shown in the plane of $\phi$ and $k$. Moreover, the rotational isodiffusivity lines of hard ellipses reproduce qualitatively the shape of the I-N transition line at least for sufficiently small $D_{\theta}$ values, which is similar as the result in hard ellipsoids of revolution~\cite{DeMichele}. The latter observation indicates that the rotational motions of the ellipses is mostly controlled by the particle's anisotropy. On the other hand, the translational isodiffusivity lines in uniaxial hard ellipsoids mimic the swallowlike shape of the coexistence line between the isotropic liquid and the crystalline phase~\cite{DeMichele}, suggesting a non-monotonic change of $D_{T}$ with $k$ at fixed density (see Fig. 2 in Ref.~\cite{DeMichele}). Note that the swallowlike shape appears because there is a distinction between prolate (rodlike) and oblate (disklike) particles for hard ellipsoids, while such a distinction does not exist in hard ellipses since there are only two symmetry axes. Our results, already apparent from Fig. 15, show that $D_{T}$ monotonically increases with $k$ at fixed $\phi$, which remains valid even when a phase transition intervenes. Thus, it appears that the spatial dimension does influence the translational dynamics. However, it should be noted that our result does not rule out a non-monotonic dependence of $D_{T}$ on $k$ at even larger fixed densities than those presented in Fig. 17. In fact, our preliminary study for polydisperse hard ellipses (which prevent the formation of crystal phases) does show that the translational glassy dynamics depends on aspect ratio non-monotonically at large fixed densities. Therefore, a common mechanism will be responsible for the translational glassy dynamics of hard ellipsoidal particles although different origins may exist at low densities.

\section{Conclusions}

In summary, we have presented a numerical study of hard ellipses over a wide range of aspect ratios and area fractions, covering state points including the isotropic, plastic and nematic phases. We have provided the results of the EOS for a series of $k$ values and compared the simulations with a recent prediction of the SPT in the isotropic phase. We find that SPT well describes the isotropic branch of the EOS in hard ellipses. Dependence of $k$ on the reduced pressure $p$ has been explored in the various phases. At large fixed area fractions, $p$ increases with $k$ in both the isotropic and the plastic phase, and interestingly, its dependence on $k$ is rather weak in the nematic phase. We rationalize the thermodynamics of hard ellipses in terms of particle motions, which show different features in different phases. For the static structures, our simulations show that the plastic crystal forms for aspect ratios up to $k=1.4$, while appearance of the stable nematic phase starts approximately at $k=3$. The locations of the I-P transition and the I-N transition have been determined respectively by analyzing the bond-orientation correlation functions and the angular correlation functions and compared with previous simulations. We demonstrate that the I-P transition point increases with $k$ as the particle's anisotropy tends to destroy the positional order, while a larger $k$ leads to a smaller area fraction where the I-N transition takes place, revealing the fact that a system of more elongated hard ellipses has a stronger tendency to form nematic liquid crystals. In addition, our results strongly support that the 2D nematic phase in hard ellipses has only quasi-LRO, as evidenced by the algebraic decay of the angular correlation function in the nematic phase. We have investigated the self-diffusion of hard ellipses and found that a phase transition in the translational degrees of freedom is indeed reflected by a corresponding change in the rotational diffusion and vice versa. Hence, we reveal a clear connection between the structure and the self-diffusion. We discuss the relevance of our results to the glass transition in hard ellipses and infer that there is an intersection point between the translational and the rotational glass transition lines in the phase diagram of glassy hard ellipses. Finally, we have evaluated the isodiffusivity lines for hard ellipses and discussed the effect of spatial dimension on the translational dynamics by comparing the diffusive dynamics of hard ellipses with that of hard ellipsoids. Our results are also valuable for understanding the structure and the diffusion of anisotropic molecules at membranes and interfaces.

\begin{acknowledgments}
This work is subsidized by the National Basic Research Program of China (973 Program, 2012CB821500), and supported by the National Natural Science Foundation of China (21222407, 21074137, 50930001) programs.
\end{acknowledgments}

\appendix

\section{Event-driven molecular dynamics for hard ellipses}

We describe the main ingredients for implementing an EDMD simulation of hard ellipses in this section. In order to perform a MD or a MC simulation for a hard-particle system, a key task is to detect overlap between two particles. This is a trivial thing for spherical particles, while it is highly non-trivial for non-spherical particles. In the case of hard ellipses, Vieillard-Baron~\cite{VieillardBaron} first introduced an overlap criterion about four decades ago. Later on, Perram and Wertheim~\cite{PW} derived a contact function. Both criteria can be easily applied in a numerical simulation of hard ellipses. However, it appears that the Perram-Wertheim (PW) approach is more convenient both computationally and theoretically. Moreover, in addition to an overlap criterion, we also need the first-order time derivative of the distance of closest approach between two moving ellipses in order to determine the contact time, which is a central quantity in an EDMD simulation~\cite{Allen}. Indeed, Donev \emph{et al}~\cite{CDMD2} have derived the time derivatives of the PW contact function and developed an EDMD algorithm for hard ellipsoidal particles. We thus follow their results and adopt the PW approach in our study. However, we note that other approaches also exist for detecting overlap between two ellipses~\cite{Overlap1, Overlap2, Overlap3}.

Let us first describe the orientation of a rigid body. In two dimensions, it needs only one angle to represent the orientation of the rigid body as there is only one rational degree of freedom. Let $\theta$ denote the angle between the semi-major axis of the ellipse and the $x$ axis, then the rotation matrix reads
\begin{equation}
\mathbf{Q}=\begin{bmatrix}\cos \theta&\sin \theta\\ -\sin \theta&\cos \theta\end{bmatrix}.
\end{equation}
Taking its shape into account, an ellipse can be described by
\begin{equation}
\mathbf{X}=\mathbf{Q}^{T}(\mathbf{O}^{-1})^{2}\mathbf{Q},
\end{equation}
where $\mathbf{O}$ is a diagonal matrix containing the semi-major and semi-minor axes along the diagonal, $\mathbf{O}^{-1}$ the inverse of $\mathbf{O}$ and $\mathbf{Q}^{T}$ the transpose of $\mathbf{Q}$.

We define the PW contact function for two ellipses $A$ and $B$ as~\cite{PW}
\begin{equation}
f_{AB}(\lambda)=\lambda(1-\lambda)\mathbf{r}_{AB}^{T}\mathbf{Y}^{-1}\mathbf{r}_{AB},
\end{equation}
where $\mathbf{r}_{AB}=\mathbf{r}_{B}-\mathbf{r}_{A}$ with $\mathbf{r}_{A}$ the position of ellipse $A$, and $\mathbf{Y}=\lambda X_{B}^{-1}+(1-\lambda)X_{A}^{-1}$. Perram and Wertheim~\cite{PW} has proven that $f_{AB}(\lambda)$ is strictly concave on the interval $[0, 1]$ and has a unique maximum $F_{AB}$ at $\lambda=\Lambda$. Then the numerically determined maximum $F_{AB}$ has the following properties:
\begin{equation}
\begin{cases}
F_{AB}>1 & \text{if $A$ and $B$ do not overlap,}\\
F_{AB}=1 & \text{if $A$ and $B$ are externally tangent,}\\
F_{AB}<1 & \text{if $A$ and $B$ overlap.}
\end{cases}
\end{equation}
Note that $f_{AB}(\lambda)$ is indeed a rational function of $\lambda$ and hence its maximum can be readily found using a Newton-Raphson method~\cite{Recipe}. When $F_{AB}$ and $\Lambda$ are known, we can scale the two ellipses by a common factor~\cite{CDMD2} so that they are externally tangent. The contact point is then determined by
\begin{equation}
\mathbf{r}_{C}=\mathbf{r}_{A}+(1-\Lambda)X_{A}^{-1}\mathbf{n}=\mathbf{r}_{B}-\Lambda X_{B}^{-1}\mathbf{n},
\end{equation}
where $\mathbf{n}=\mathbf{Y}^{-1}\mathbf{r}_{AB}$ is the unnormalized common normal vector at the contact point. Then $\mathbf{r}_{C}$ and $\mathbf{n}$ are used to compute the first-order time derivative of $F_{AB}$, which has been derived in Ref.~\cite{CDMD2} and reads
\begin{equation}
\dot{F}_{AB}=2\Lambda(1-\Lambda)\mathbf{n}^{T}\mathbf{v}_{C},
\end{equation}
where $\mathbf{v}_{C}$ is the projection of the relative velocity at the contact point
\begin{equation}
\mathbf{v}_{C}=(\mathbf{v}_{B}+\omega_{B}\boxtimes\mathbf{r}_{BC})-(\mathbf{v}_{A}+\omega_{A}\boxtimes\mathbf{r}_{AC}),
\end{equation}
where $\mathbf{r}_{BC}=\mathbf{r}_{C}-\mathbf{r}_{B}$,  $\mathbf{r}_{AC}=\mathbf{r}_{C}-\mathbf{r}_{A}$, $\mathbf{v}$ and $\omega$ are the translational and angular velocities, respectively, and the cross product $\boxtimes$ is defined as
\begin{equation}
\omega\boxtimes\mathbf{r}=\begin{bmatrix}-\omega r_{y}\\ \omega r_{x}\end{bmatrix}.
\end{equation}
In principle, the contact time of two moving ellipses can be then numerically determined by using a Newton-Raphson method~\cite{Recipe}. However, some numerical problems are probably met and some tricks must be employed. The reader can find more details in Refs.~\cite{CDMD1, CDMD2}.

When the two ellipses $A$ and $B$ become externally tangent, their new translational velocities $\mathbf{v}_{A, new}$, $\mathbf{v}_{B, new}$, and angular velocities $\omega_{A, new}$ and $\omega_{B, new}$ can be exactly found from the conservation of total energy, linear and angular momentum. Specifically, let $\mathbf{v}_{A, old}$, $\mathbf{v}_{B, old}$, $\omega_{A, old}$ and $\omega_{B, old}$ denote the translational and angular velocities of ellipses $A$ and $B$ before collision, their new velocities are updated by the following formula:
\begin{equation}
\mathbf{v}_{A, new}=\mathbf{v}_{A, old}-\frac{\Delta h}{m} \mathbf{\hat{n}},
\end{equation}
\begin{equation}
\mathbf{v}_{B, new}=\mathbf{v}_{B, old}+\frac{\Delta h}{m} \mathbf{\hat{n}},
\end{equation}
\begin{equation}
\omega_{A, new}=\omega_{A, old}+\frac{\Delta h}{I}(\mathbf{r}_{AC}\times\mathbf{\hat{n}}),
\end{equation}
\begin{equation}
\omega_{B, new}=\omega_{B, old}-\frac{\Delta h}{I}(\mathbf{r}_{BC}\times\mathbf{\hat{n}}),
\end{equation}
where $m$ is the mass of the particle, $I$ is the moment of inertia, $\mathbf{\hat{n}}$ is the normalized common normal vector at the contact point, and $\Delta h$ is the momentum exchange during the collision, which is calculated according to
\begin{equation}
\Delta h=2\mathbf{\hat{n}}^T\mathbf{v}_{C}(\frac{2}{m}+\frac{|\mathbf{r}_{AC}\times\mathbf{\hat{n}}|^2}{I}+\frac{|\mathbf{r}_{AB}\times\mathbf{\hat{n}}|^2}{I})^{-1}.
\end{equation}
Meanwhile, the reduced pressure can be computed from the momentum exchange during a time interval $\Delta t$:
\begin{equation}
p=\frac{PV}{Nk_BT}=1-\sum_{\Delta t}\frac{\Delta h(\mathbf{r}_{AB}\mathbf{\hat{n}}^T)}{2NT\Delta t}.
\end{equation}

To accelerate the simulation, we use a neighbor list (NL) method. Note that the NL method for hard ellipses is more complicated than that for hard spheres~\cite{CDMD1, CDMD2}, but the essence remains unchanged.

\section{Effect of the system size}

\begin{figure}[tb]
 \centering
 \includegraphics[angle=0,width=0.45\textwidth]{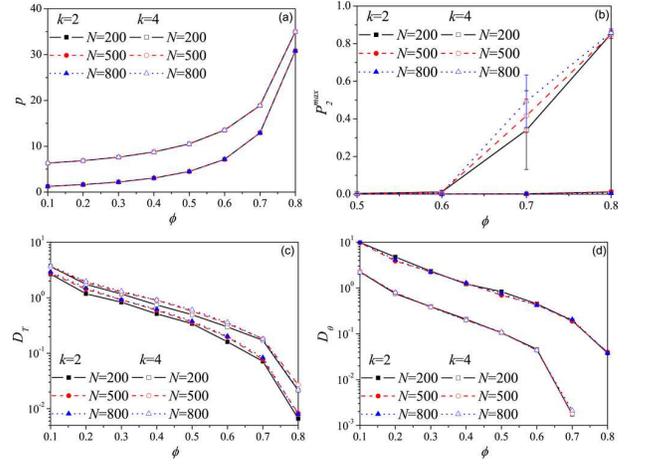}
 \caption{For hard ellipses with $k=2$ and $4$, effect of the system size $N$ on (a) reduced pressure, (b) nematic order parameter, (c) translational diffusion constant and (d) rotational diffusion constant. The results for $k=4$ in (a) are shifted up by $5$ for clarity, and the error bars in (b) correspond to the standard deviation over four independent samples. The system shows an isotropic phase in the whole density range in the case of $k=2$, while the I-N transition occurs at $\phi\approx0.7$ for $k=4$.}
\end{figure}

\begin{figure}[b]
 \centering
 \includegraphics[angle=0,width=0.45\textwidth]{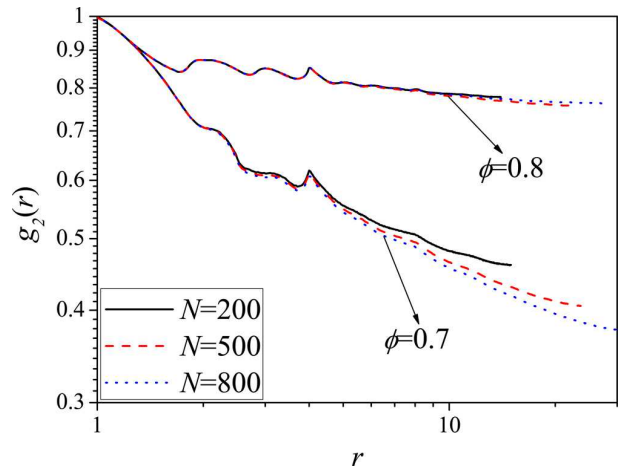}
 \caption{Effect of the system size $N$ on the angular correlation function $g_{2}(r)$ for $k=4$ at $\phi=0.7$ and $0.8$.}
\end{figure}

In this section, we assess the effect of the system size. As introduced in Sec. I, an analysis of the system-size dependence of the orientational order parameter is very useful especially in the 2D nematic phase. However, we also want to explore the influence of the system size on other quantities in order to confirm the validity of the conclusions given in this work. To this end, we also performed simulations of hard ellipses with $N=200$ and $800$ for $k=2$ and $4$. We have concluded in Sec. II that the system shows an isotropic phase at area fractions up to $\phi=0.8$ in the case of $k=2$ during the compression run, while the isotropic liquid transforms into the nematic liquid crystal at $\phi\approx0.7$ for $k=4$. We find from Fig. 18 that the system size does not affect the above conclusion. In addition, all properties, including the reduced pressure $p$, the nematic order parameter $P_{2}^{max}$ and the diffusion constants $D$, do not depend on the system size within the statistical accuracy at least for the studied particle numbers. In particular, $P_{2}^{max}$ is also independent of $N$ even in the nematic phase of hard ellipses. This immediately indicates that the computed $P_{2}^{max}$ here is different from the order parameter $q$ obtained from the tensor order parameter, since previous simulations do show that $q$ decreases algebraically with $N$~\cite{Frenkel1}. Thus, we cannot determine the point of the I-N transition from the dependence of the system size on $P_{2}^{max}$, although we have shown that $P_{2}^{max}$ is more sensitive to the onset of the orientational order than $q$ at fixed particle number. Moreover, as shown in Fig. 19, the angular correlation function $g_{2}(r)$ decays faster for larger $N$ in the nematic phase, but this feature seems to become less evident and even disappears as $\phi$ gets larger. However, the results confirm again that the nematic phase in hard ellipses does have only quasi-LRO.


\begin{thebibliography}{0}
\bibitem{footnote1}Note that the free energy density for a nematic liquid crystal in three dimensions is composed of splay, twist and bend energy terms and that each term is associated with a separate elastic constant, while the twist term is absent in two dimensions. Since the three elastic constants are of the same order of magnitude in most cases, the free energy is often derived by assuming that the three elastic constants are equal. Based on this so-called one-constant approximation, the resulting free energy has a very compact form and thus facilitates the theoretical analysis~\cite{DeGennes}.
\bibitem{Comment}Note that even if the term ``bond-orientation'' is used, the calculation of the bond-orientation correlation function is based solely on the centers of mass of the particles and thus it indeed measures the positional order.
\end{thebibliography}

\end{document}